\input phyzzx
\input epsf
\def\INSERTFIG#1#2#3{\epsfysize=#1in
	\hbox to\hsize{\hfil\epsffile{#2}\hfil} #3}

\hoffset=0.2truein
\voffset=0.1truein
\hsize=6truein
\def\TITLEPAGE{\frontpagetrue}
\def\CMUHEP#1{\hbox to\hsize{\twelvepoint \baselineskip=12pt
        \hfil\vtop{
        \hbox{\strut CMU-HEP#1}
	\hbox{\strut DOE-ER/40682-119}}}}
\def\CMU{\medskip
	\centerline{\it Department of Physics, Carnegie Mellon University,}
	\centerline{\it Pittsburgh, PA 15213}}
\def\TITLE#1{\vskip .5in \centerline{\bf \fourteenpoint #1}}
\def\AUTHOR#1{\vskip .2in \centerline{#1}}

\def\ABSTRACT#1{\vfill\vskip .2in \centerline{\twelvepoint \bf Abstract}
        #1}
\def\DATE#1{\vfill\noindent #1}
\def\ENDTITLEPAGE{\vfil\eject\pageno=1}

\TITLEPAGE
\CMUHEP{96-08}
\TITLE{Shape Variables for the Doubly-Charmed B Decays}
\AUTHOR{Ming Lu\footnote{\dag}{Email address: lu@fermi.phys.cmu.edu}}
\CMU
\ABSTRACT{We propose to use the hadronic shape variables to test the
assumption of ``local duality'' in the doubly-charmed B decays,
$B\rightarrow X_{c\bar c s}$.
We compute these shape variables at leading order in parton model,
and discuss how they can be compared with the experimental values.
}

\DATE{June, 1996}
\ENDTITLEPAGE

Recently there has been a lot of progress in understanding inclusive B decays
using the operator product expansion (OPE) and the heavy quark effective theory
[1].
A better understanding of inclusive B decays will enable us to calculate
the semileptonic and nonleptonic decay rates with less theoretical
uncertainties, and therefore help us to get a more
precise determination of the CKM matrix elements $|V_{cb}|$ and $|V_{ub}|$.
In addition, the B semileptonic and nonleptonic decay rates also determine
the semileptonic branching fraction of B decays\footnote{1}{At one point 
it was thought that the theoretical value for the
semileptonic branching fraction is too high to be consistent with the
experimental value [2]. However, given the uncertainties associated with the 
theoretical prediction, it seems
premature to talk about its discrepancy with the experimental value [3].}.

The decay rates of B into inclusive semileptonic and nonleptonic final states
have an OPE, of which the leading order reproduces the parton model result.
However, the validity of the OPE is different for semileptonic
and nonleptonic decays.
For inclusive semileptonic B decays ($b\rightarrow c\ell \bar\nu$ at quark
level),
the decay rate (and the differential spectra)
can be expressed as an integral over some kinematic variable.
In the complex plane of this kinematic variable,
the only singularity is the physical cut on the real axis.
It is possible to deform the integration contour for the decay rate (and the
differential spectra) of this
kinematic variable away from the physical cut in the complex plane.
Along the deformed contour away from the physical cut,
one hopes that the resonance effects associated with long distance hadronic
production can be avoided, and perturbative QCD can be used reliably to compute
the inclusive decay rate
(and the differential spectra). This is called ``global duality".
It is generally believed to be valid.
For inclusive nonleptonic B decays ($b\rightarrow c \bar u d$,
$b\rightarrow c\bar c s$ at quark level), however, one has to invoke
quark-hadron duality directly in the physical region in order to use
the OPE (and perturbative QCD) to compute
the decay rates. This assumption is called ``local duality".
The validity of local duality relies on the energy release in the B decay
being much larger than $\Lambda_{QCD}$, in which case there are many available 
states in the decay products. It is then likely that the decay
rate is insensitive to the resonance effects associated with the production 
of particular
hadron state, and one may apply local duality in the physical region
to compute the B decay rate.
However, it is not clear whether the B meson is massive enough for local
duality to apply.
Especially in the case of the doubly-charmed B decays ($b\rightarrow c\bar c s$
at quark level),
due to the large mass of the charm quark,
the energy release $\Delta = m_b - 2 m_c$ divided among three partons in the
final state is not much larger
than the QCD scale, and the resonance effects
may be significant enough to destroy the validity of local duality. Therefore,
it seems particularly
important to test the validity of local duality for inclusive nonleptonic B
decays in as many ways as possible,
especially for the doubly-charmed decays. In this letter, we propose to use the
shape variables
of the doubly-charmed B decays as a test of local duality.

The shape variables were first introduced in [4] for B decays into semileptonic
final states and
singly-charmed nonleptonic final states. In analogy we can define
a similar shape variable for B decays into doubly-charmed final states
($b\rightarrow c\bar c s$)
$$
{d\Sigma \over d\cos\theta} = {1\over \Gamma} \Biggl[\sum_{i,j} \int dE_i dE_j
	{|\vec p_i| \over m_b} {|\vec p_j| \over m_b} {d^3\Gamma\over dE_i dE_j
d\cos\theta} +
$$
$$
	\sum_i \int dE_i {|\vec p_i|^2 \over m_b^2} {d\Gamma\over dE_i} \delta
(1-\cos\theta)\Biggr]~~.
\eqno (1)
$$
$\Gamma$ is the total decay width of B into doubly-charmed final states, $m_b$
is the b-quark mass. The first
sum is over any hadronic pair $(i,j)$ for $i\neq j$, with $(i,j)$ and $(j,i)$
counted as different pairs.
$\theta$ is the angle between the two hadrons in $(i,j)$ pair. As noted in [4],
the shape variable
defined above is infrared safe, free from divergences due to collinear gluon
and soft gluon emissions.
Consequently one can use free quarks and gluons to calculate this shape
variable, i.e., one can compute the differential decay rate at parton level,
and sum over all possible parton pairs (and partons)
to get ${d\Sigma \over d\cos\theta}$. Notice that for this
observable there is no OPE, unlike for the inclusive decay rates. All that is
being assumed is local duality.

In the limit $m_b \gg m_c \gg \Lambda_{QCD}$, the $c$ and the $\bar c$ in the
decay products move along
different directions with momenta much larger than
$\Lambda_{QCD}$ (in $b$ rest frame) in most of the phase space.
Because $c$($\bar c$) quark is heavy, the effect of long-distance soft QCD
interaction on the direction of
$c$($\bar c$) is suppressed.
To leading order in ${\cal O}(\Lambda_{QCD}/m_c)$, the direction of $c$($\bar
c$) quark is a
physical observable.
One can define a shape variable for production of charmed and anti-charmed
final states only
$$
{d\Sigma_c \over d\cos\theta} = {1\over \Gamma} \sum_{c,\bar c} \int dE_c
dE_{\bar c}
	{|\vec p_c| \over m_b} {|\vec p_{\bar c}| \over m_b} {d^3\Gamma\over dE_c
dE_{\bar c} d\cos\theta}~~,
	\eqno (2)
$$
where the summation is over all possible charmed and anti-charmed hadrons,
$\theta$ is the angle between
the charmed hadron and the anti-charmed hadron. As noted above, one can invoke
local duality to calculate this
shape variable using the free quark picture, and it should be valid to leading
order in
${\cal O}(\Lambda_{QCD}/m_c)$. Notice that
it is not clear at which order the QCD nonperturbative corrections
arise for the parton model result of ${d\Sigma \over d\cos\theta}$ [5].
Consequently the parton model result for ${d\Sigma_c \over d\cos\theta}$
ought to be more reliable.

The shape variables ${d\Sigma \over d\cos\theta}$ and 
${d\Sigma_c \over d\cos\theta}$ are
computed in leading order in parton model for various values of $m_c/m_b$.
The results are shown in fig.~1. (We have taken $m_s = 0$. 
${d\Sigma \over d\cos\theta}$
for $m_c = 0$, which is for decay $b\rightarrow u\bar u d$,
 was first computed in [4].) Due to kinematics, the shape
variables are all strongly peaked
at $\cos\theta = -1$, and fall off rapidly as $\cos\theta$ goes from $-1$ to
$+1$.
The falloff is more steep for  ${d\Sigma_c \over d\cos\theta}$,
which implies that the hadrons containing $c$ and $\bar c$ respectively in the
decay products prefer to travel
back to back. It will be interesting to see whether the experimental data
reproduce this feature.
The $\delta$-function contribution to ${d\Sigma \over d\cos\theta}$
at $\cos\theta = +1$ is shown in Fig.~2
as a function of $m_c/m_b$.

Local duality has been used to compute these shape variables. To compare with
the experimental measurements,
one has to smear both the calculated and the measured shape variables [6]. The
particular details of the
smearing function are not important, except for its width. 
Take ${d\Sigma_c \over d\cos\theta}$ for example. Consider the contributions 
to it at hadronic level.
There is a significant branching fraction for producing charmonium ($c\bar c$)
final states. Its contribution to ${d\Sigma_c \over d\cos\theta}$ is a $\delta$-function at $\cos\theta = +1$. This
$\delta$-function is not found in the free quark picture, as long-distance
nonperturbative QCD effects
are essential in forming $c\bar c$ bound states. There is also a large 
branching fraction for $D\bar D$ two-body final
states, which give a $\delta$-function contribution at $\cos\theta = -1$ in 
${d\Sigma_c \over d\cos\theta}$. In the free quark picture 
${d\Sigma_c \over d\cos\theta}$ is only strongly peaked at $\cos\theta = -1$.
Therefore smearing is crucial in comparing the theoretically-calculated shapes
with the experimentally-measured ones.

It is useful to define the Fox-Wolfram moments [7]
$$
H(n) = \int d\cos\theta {d\Sigma \over d\cos\theta} P_n(\cos\theta)~~,
$$
$$
H_c(n) = \int d\cos\theta {d\Sigma_c \over d\cos\theta} P_n(\cos\theta)~~,
{}~~~\eqno (3)
$$
where $P_n(x)$ are Legendre polynomials.
The first several moments are presented in Tables~1 and 2. 
Notice that $H(1) = 0$ because of momentum conservation.
In Table~1 $H(0)$ is subtracted from all the moments
to remove the $\delta$-function contribution to ${d\Sigma \over d\cos\theta}$.
Apart from the contribution from the $\delta$-function in ${d\Sigma \over
d\cos\theta}$, the high moments go to zero asymptotically as $n\rightarrow
\infty$. These moments can be measured experimentally, which will test
the applicability of local duality to $B\rightarrow X_{c\bar c s}$.

\goodbreak \midinsert
\centerline{ \vbox{\offinterlineskip
 \halign to3.8in{\vrule#\tabskip=2em plus1em minus.5em&		
   \hfil $#$ \hfil & \vrule# &					
   \hfil $#$ \hfil & \vrule# &					
   \hfil $#$ \hfil & \tabskip=0pt\vrule#\cr			
 \noalign{\hrule}						
 height14pt depth5.5pt& n && \multispan3 \hfil$H(n)-H(0)$\hfil &\cr
 && \multispan5 \smash{\vrule height.4pt width222pt depth0pt} \cr
 height12pt depth7.5pt&   && m_c/m_b=0.3 && m_c/m_b=0.37 &\cr
 \noalign{\hrule}
 height14pt depth5.5pt& 1 && -0.49 && -0.30 &\cr
 height12pt depth5.5pt& 2 && -0.22 && -0.13 &\cr
 height12pt depth5.5pt& 3 && -0.36 && -0.22 &\cr
 height12pt depth5.5pt& 4 && -0.29 && -0.17 &\cr
 height12pt depth5.5pt& 5 && -0.33 && -0.20 &\cr
 height12pt depth5.5pt& 6 && -0.30 && -0.18 &\cr
 \noalign{\hrule}
 } } }
\bigskip \centerline{Table 1. Fox-Wolfram moments for 
${d\Sigma \over d\cos\theta}$.} \endinsert

\medskip
\goodbreak \midinsert
\centerline{ \vbox{\offinterlineskip
 \halign to3.8in{\vrule#\tabskip=2em plus1em minus.5em&		
   \hfil $#$ \hfil & \vrule# &					
   \hfil $#$ \hfil & \vrule# &					
   \hfil $#$ \hfil & \tabskip=0pt\vrule#\cr			
 \noalign{\hrule}						
 height14pt depth5.5pt& n && \multispan3 $H_c(n)$ &\cr
 && \multispan5 \smash{\vrule height.4pt width222pt depth0pt} \cr
 height12pt depth7.5pt&   && m_c/m_b=0.3 && m_c/m_b=0.37 &\cr
 \noalign{\hrule}
 height14pt depth5.5pt& 0 && 5.8\times 10^{-2} && 3.8\times 10^{-2} &\cr
 height12pt depth5.5pt& 1 && -3.9\times 10^{-2} && -2.9\times 10^{-2} &\cr
 height12pt depth5.5pt& 2 && 2.4\times 10^{-2} && 2.0\times 10^{-2} &\cr
 height12pt depth5.5pt& 3 && -1.4\times 10^{-2} && -1.3\times 10^{-2} &\cr
 height12pt depth5.5pt& 4 && 8.7\times 10^{-3} && 8.9\times 10^{-3} &\cr
 height12pt depth5.5pt& 5 && -5.5\times 10^{-3} && -6.0\times 10^{-3} &\cr
 \noalign{\hrule}
 } } }
\bigskip \centerline{Table 2. Fox-Wolfram moments for 
${d\Sigma_c \over d\cos\theta}$.} \endinsert

New physics other than the Standard Model of electroweak interactions
can have different Lorentz structures for the $b$ decay amplitude,
such as scalar-scalar type or $(V+A)$ type.
It turns out that these various forms of the decay amplitudes
all produce the similar plots
of the shape variables at leading order in parton model. 
For example, the first several Fox-Wolfram moments $H_c(n)$
from the $(V+A)\times (V-A)$ type Lorentz structure
differ from the Standard Model results by less than $10\%$ at leading
order in parton model.
So the shape variables are not sensitive probes for new physics.
The experimental measurements of these shape variables will constitute a test
of local duality, rather than
a test for the possibility of new physics.

In summary, we have computed the shape variables for the doubly-charmed
inclusive B decays.
The results found here for B decays also apply for $B_s$ and $\Lambda_b$
decays.
Although we have only computed the leading order parton model
contributions, it is straightforward to include the ${\cal O}(\alpha_s)$
perturbative contributions.
Measurements of these shape variables will constitute a test of
the applicability of local duality in the inclusive B decays.
As there is no OPE for these shape variables,
it is possible that local duality is more likely to fail for these shape
variables than for the decay rates.
It should be realized that understanding of the nonperturbative corrections
to these shape variables
is poor. If the experimental measurements of these shape variables corroborate
the theoretical calculations, then it is possible that
local duality holds in the decay $B\rightarrow  X_{c\bar c s}$
despite the small energy release, and we will be
more confident in using it to calculate the inclusive decay rate.

\bigskip

\centerline {\bf Acknowledgement}
I am grateful to Martin Savage for suggestion to look at the hadronic event
shapes and many useful discussions. I also thank Zoltan Ligeti and 
Lincoln Wolfenstein for many interesting comments.
This work was supported in part by DOE under contract DE-FG02-91ER40682.

\bigskip\bigskip

\centerline {\bf References}

\item{1.} M.~Voloshin and M.~Shifman, Sov. J. Nucl. Phys. {\bf 41} (1985)120;
J.~Chay, H.~Georgi, and
B.~Grinstein, Phys. Lett. {\bf B247} (1990)399; A.V.~Manohar and M.B.~Wise,
Phys. Rev. {\bf D49} (1994)1310;
T.~Mannel, Nucl. Phys. {\bf B413} (1994)396; I.I.~Bigi, N.G.~Uraltsev, and
A.I.~Vainshtein, Phys. Lett. {\bf
B293} (1992)430; I.I.~Bigi, M.~Shifman, N.G.~Uraltsev, and A.I.~Vainshtein,
Phys. Rev. Lett. {\bf 71} (1993)496;
B.~Blok, L.~Koyrakh, M.~Shifman, and A.I.~Vainshtein, Phys. Rev. {\bf D49}
(1994)3356; Erratum, Phys. Rev. {\bf D50} (1994)3572.

\item{2.} G.~Altarelli and S.~Petrarca, Phys. Lett. {\bf B261} (1991)303;
I.~Bigi, B.~Blok, M.~Shifman, and A.~Vainshtein, Phys. Lett. {\bf B323} 
(1994)408. See also
A.F.~Falk, M.B.~Wise, and I.~Dunietz, Phys. Rev. {\bf D51} (1995)1183.

\item{3.} M.~Neubert, hep-ph/9604412; 
M.~Lu, M.~Luke, M.~Savage, and B.~Smith, hep-ph/9605406.

\item{4.} M.~Luke, M.J.~Savage, and M.B.~Wise, Phys. Lett. {\bf B322}
(1994)154.

\item{5.} See, e.g., A.V.~Manohar and M.B.~Wise, Phys. Lett. {\bf B344}
(1995)407.

\item{6.} E.C.~Poggio, H.R.~Quinn, and S. Weinberg, Phys. Rev. {\bf D13}
(1976)1958.

\item{7.} G.C.~Fox and S.~Wolfram, Phys. Rev. Lett. {\bf 41} (1978)1581.

\vfill\eject

\INSERTFIG{3.5}{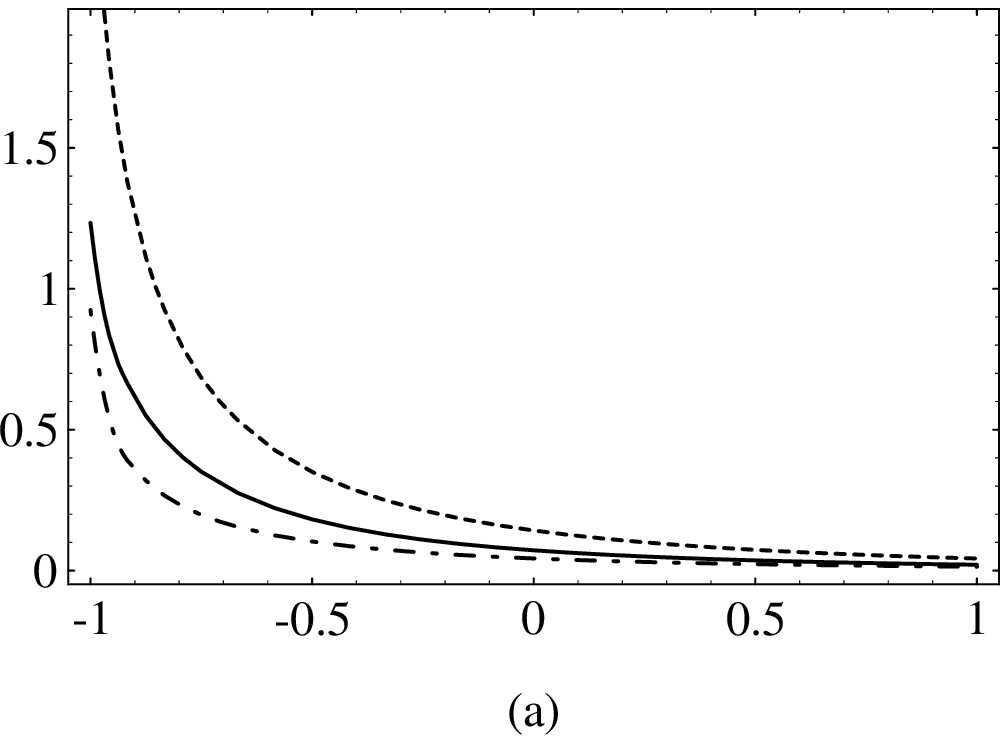}{}
\INSERTFIG{3.5}{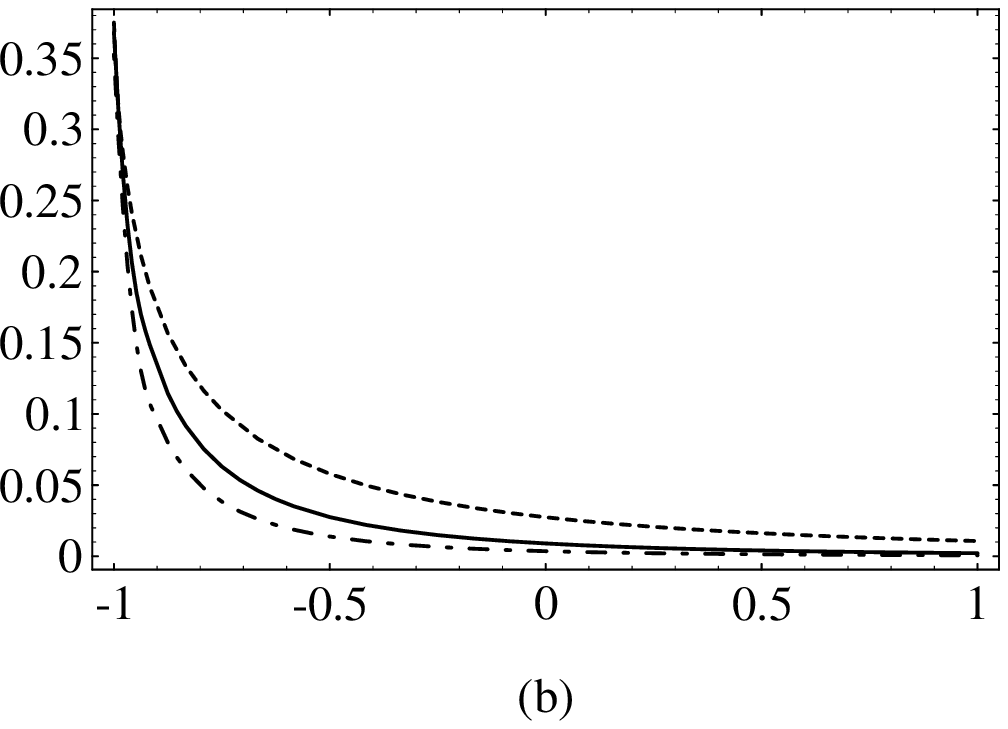}{Fig.~1: Shape variables (a) ${d\Sigma \over 
d\cos\theta}$ and (b) ${d\Sigma_c \over d\cos\theta}$ plotted over 
$\cos\theta$. The dashed line is for $m_c/m_b = 0$, the solid line 
$m_c/m_b = 0.30$, and the dot-dashed line $m_c/m_b = 0.37$.}

\vfill\eject

\INSERTFIG{3.5}{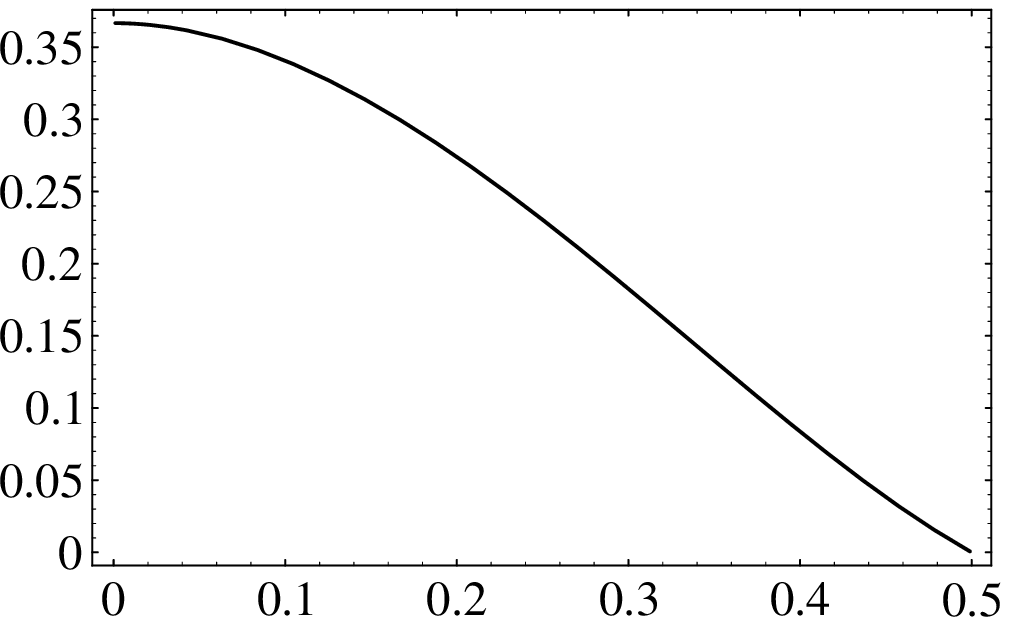}{Fig.~2: The coefficient of $\delta(1-\cos\theta)$
term in ${d\Sigma \over d\cos\theta}$ plotted over $m_c/m_b$.}

\vfill\eject

\end